\documentstyle[prl,aps,twocolumn,epsf]{revtex}

\newcommand{\be}{\begin{equation}}
\newcommand{\ee}{\end{equation}}
\newcommand{\bea}{\begin{eqnarray}}
\newcommand{\eea}{\end{eqnarray}}

\newcommand{\aeq}{&=&}

\newcommand{\itDelta}{{\it \Delta}}

\newcommand{\itPi}{{\it \Pi}}

\newcommand{\bra}{\langle}
\newcommand{\ket}{\rangle}
\newcommand{\dbra}{\bra \! \bra}
\newcommand{\dket}{\ket \! \ket}

\newcommand{\me}{\mbox{e}}

\newcommand{\bq}{{\bar q}}
\newcommand{\rK}{{\rm K}}
\newcommand{\rT}{{\rm T}}

\newcommand{\rR}{{\rm R}}

\newcommand{\rRe}{{\rm Re}}

\newcommand{\rN}{{\rm N}}
\newcommand{\rS}{{\rm S}}

\begin{document}


\draft

\title{Multifractal Analysis of Turbulence by Statistics
based on Non-Extensive Tsallis' or Extensive R\'{e}nyi's 
Entropy
}

\author{N.~Arimitsu
}
\address{Graduate School of EIS, Yokohama Nat'l.~University, 
Yokohama 240-8501, Japan 
}
\author{T.~Arimitsu
}
\address{Institute of Physics, University of Tsukuba,
Ibaraki 305-8571, Japan 
}

\date{September 7, 2001}

\maketitle

\begin{abstract}

An analytical expression of probability density function (PDF) of
velocity fluctuation is derived with the help of the statistics
based on generalized entropy (the Tsallis entropy or the R\'{e}nyi entropy).
It is revealed that the derived PDF explains the detailed structure
of experimentally observed PDF as well as the scaling exponents of 
velocity structure function. 
Every parameters appeared in the analysis, including
the index proper to the Tsallis entropy or the R\'{e}nyi entropy, are 
determined, self-consistently, by making use of observed value of 
intermittency exponent. 
The experiments conducted by Lewis and Swinney (1999) are analyzed.

\end{abstract}


\narrowtext



An investigation of turbulence based on the generalized entropy, 
Tsallis' \cite{Tsallis88,Tsallis99}
or R\'{e}nyi's \cite{Renyi}, was
started by the present authors~\cite{AA} with an investigation 
of the p-model~\cite{Meneveau87a,Meneveau87b} in terms of 
a generalized statistics constructed on the entropy.
We intend to analyze the velocity fluctuation, 
$
\delta u(r) = \vert u(x+r) - u(x) \vert
$,
between two points with a distance $r$ apart 
in a turbulence. Here, $u$ represents one of the components 
of the fluid velocity field ${\vec u}$, say $x$-component.
By making use of the velocity fluctuation $\delta u_0$ of eddies
with the largest size $\ell_0$,
the Reynolds number is estimated, in the absence of intermittency
\cite{K41}, as 
$
{\rm Re} = \delta u_0 \ell_0/\nu = ( \ell_0/\eta )^{4/3}
$
where $\nu$ and 
$
\eta = ( \nu^3/\epsilon )^{1/4}
$
are, respectively, the kinematic viscosity and 
the Kolmogorov scale.
The quantity $\epsilon$ represents the energy input rate at the largest eddies.

In the case of high Reynolds number $\rRe \gg 1$, there exist 
a lot of steps, $n=1,2,\cdots$, at each of which eddies break up into 
two parts producing a energy cascade. The size of eddies in the $n$th step
of the cascade is assumed to be given by 
$
\ell_n = \delta_n \ell_0
$
with
$\delta_n = 2^{-n}$.
Then, our main interest in the following reduces to the fluctuation
of velocity difference $\delta u_n = \delta u(\ell_n)$ corresponding to 
the size of $n$th eddies in the cascade.
Note that the dependence of the number of steps $n_{\rK}$ on $r/\eta$, 
within the analysis where intermittency is not taken into account \cite{K41}, 
is given by
\be
n = - \log_2 r/\eta + (3/4) \log_2 \rRe.
\label{n-roeta no intermittency}
\ee

In this paper, we examine the experimental results obtained by 
Lewis and Swinney \cite{Lewis-Swinney99} for turbulent Couette-Taylor flow
produced in a concentric cylinder system.
Our analysis is based on the fact that, for high Reynolds number
$\rRe \gg 1$, the Navier-Stokes equation
for incompressible fluid is invariant under 
the scale transformation~\cite{Meneveau87b}:
${\vec r} \rightarrow \lambda^{\alpha/3} {\vec u}$, 
${\vec u} \rightarrow \lambda {\vec r}$, 
$t \rightarrow \lambda^{1- \alpha/3} t$ and 
$\left(p/\rho\right) \rightarrow \lambda^{2\alpha/3} \left(p/\rho\right)$.
Here, the exponent $\alpha$ is an arbitrary real quantity which specifies the degrees
of singularity in the velocity gradient
$
\left\vert \partial u(x)/\partial x \right\vert 
= \lim_{\ell_n \rightarrow 0} \vert u(x+\ell_n) - u(x) \vert/\ell_n
= \lim_{\ell_n \rightarrow 0} \delta u_n/\ell_n
$.
This can be seen with the relation
$
\delta u_n / \delta u_0 = (\ell_n / \ell_0)^{\alpha/3},
\label{u-alpha}
$
which leads to the singularity in the velocity gradient~\cite{Benzi84}
for $\alpha < 3$, since 
$
\delta u_n/\ell_n^{\alpha/3} = {\rm const.}
$.

The distribution 
$
P^{(n)}(\alpha) d\alpha \propto P^{(1)}(\alpha)^n d\alpha 
$
of singularities at the $n$th step in the cascade with 
\be
P^{(1)}(\alpha) \propto \left[ 1 - (\alpha - \alpha_0)^2 \big/ (\itDelta \alpha )^2 
\right]^{1/(1-q)},
\label{Tsallis prob density}
\ee
$
(\itDelta \alpha)^2 = 2X \big/ [(1-q) \ln 2 ]
$
was derived \cite{AA1,AA2,AA3,AA4} by taking an extreme of 
the Tsallis entropy \cite{Tsallis88,Tsallis99,Havrda-Charvat}
$
S_{q}^{\rT}[P^{(1)}(\alpha)] = \left(1-q \right)^{-1} 
\left(\int d \alpha P^{(1)}(\alpha)^{q} -1 \right)
\label{SqTHC-alpha}
$
which is non-extensive, or the R\'{e}nyi entropy \cite{Renyi}
$
S_{q}^{\rR}[P^{(1)}(\alpha)] = \left(1-q \right)^{-1} 
\ln \int d \alpha P^{(1)}(\alpha)^{q}
\label{SqR-alpha}
$
which has the extensive character,
with appropriate constraints, i.e., the normalization of distribution function,
$
\int d\alpha P^{(1)}(\alpha) = \mbox{const.}
\label{cons of prob}
$,
and the $q$-variance being kept constant as a known quantity,
$
\sigma_q^2 = \bra (\alpha- \alpha_0)^2 \ket_{q}
= (\int d\alpha P^{(1)}(\alpha)^{q} 
(\alpha- \alpha_0 )^2 ) / \int d\alpha P^{(1)}(\alpha)^{q}
\label{q-variance}
$.
In deriving $P^{(n)}(\alpha)$, we assumed that each step in the cascade is 
statistically independent.
Note that the values of $\alpha$ are restricted within the range
$[\alpha_{\rm min},\ \alpha_{\rm max}]$, where
$\alpha_{\rm max} - \alpha_0 = \alpha_0 - \alpha_{\rm min} = \Delta \alpha$.

With the help of the relation
$
P^{(n)}(\alpha) \propto \delta_n^{1-f(\alpha)}
$ \cite{Meneveau87b},
we can extract the multi-fractal spectrum $f(\alpha)$ in the form 
\cite{AA1,AA2,AA3,AA4}:
\be
f(\alpha) = 1 + (1-q)^{-1} \log_2 \left[ 1 - \left(\alpha - \alpha_0\right)^2
\big/ \left(\Delta \alpha \right)^2 \right],
\label{Tsallis f-alpha}
\ee
which reveals how dense each singularity, labeled by $\alpha$, fills 
physical space.

\begin{figure}[htbp]
\begin{center}
\leavevmode
\epsfxsize=60mm
\epsfbox{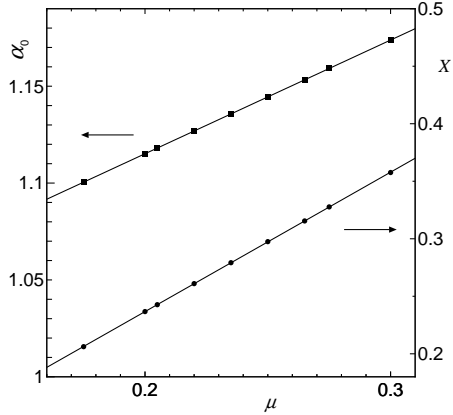}
\caption{The $\mu$-dependence of $\alpha_0$ and $X$.
Squares and circles are the points where the self-consistent equations
are solved.
}
\label{Fig: mu-dependence of alpha0 and X}
\end{center}
\end{figure}

\begin{figure}[htbp]
\begin{center}
\leavevmode
\epsfxsize=55mm
\epsfbox{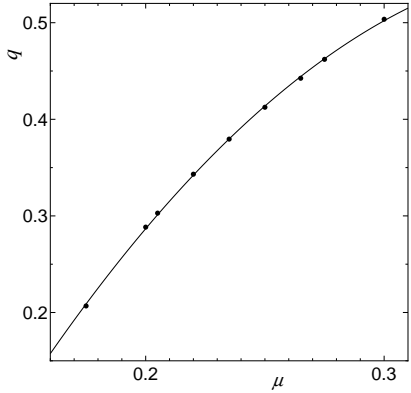}
\caption{The $\mu$-dependence of $q$.
Circles are the points where the self-consistent equations
are solved.
}
\label{Fig: mu-dependence of q}
\end{center}
\end{figure}

By making use of an observed value of the intermittency exponent $\mu$ as an input,
the quantities $\alpha_0$, $X$ and the index $q$ can be determined, 
self-consistently, with the help of the three independent equations, i.e.,
the energy conservation:
$
\left\bra \epsilon_n \right\ket = \epsilon
\label{cons of energy}
$,
the definition of the intermittency exponent $\mu$:
$
\bra \epsilon_n^2 \ket 
= \epsilon^2 \delta_n^{-\mu}
\label{def of mu}
$,
and the scaling relation
\cite{AA1,AA2,AA3,AA4}:
$
1/(1-q) = 1/\alpha_- - 1/\alpha_+
\label{scaling relation}
$
with $\alpha_\pm$ satisfying $f(\alpha_\pm) =0$,
where the average $\bra \cdots \ket$ is taken by $P^{(n)}(\alpha)$.
The scaling relation is a generalization of the one derived first by
\cite{Costa,Lyra98} to the case where the multi-fractal spectrum has 
negative values.
The $\mu$-dependences of the self-consistent solutions, $\alpha_0$, $X$ and $q$, 
are given in Fig.~\ref{Fig: mu-dependence of alpha0 and X} and 
Fig.~\ref{Fig: mu-dependence of q}
for the region where the value of $\mu$ is usually observed,
i.e., $0.16 \leq \mu \leq 0.31$.
We see that they are given in this region by the equations
$
\alpha_0 = 0.998 + 0.587 \mu
$,
$
X = - 5.73 \times 10^{-3} + 1.21 \mu
$
and
$
q = -0.607 + 6.01 \mu - 7.72 \mu^2
$.

Adopting the value $\mu = 0.28$ observed in the experiment~\cite{Lewis-Swinney99} 
with the Taylor-scale Reynolds number $R_{\lambda} = 262$
($\rRe = 540\ 000$) or $R_{\lambda} = 80$ ($\rRe = 69\ 000$), and solving 
the above three equations self-consistently, 
we have $q = 0.471$, $\alpha_0 = 1.162$, $X = 0.334$. 
Then, we obtain
$
\alpha_{+} -\alpha_0 
= \alpha_0 - \alpha_{-} = 0.748
$,
$
\itDelta \alpha = 1.350
$ 
\cite{AA1,AA2,AA3,AA4}.

It may be reasonable to assume that the probability $\itPi^{(n)}(x_n) dx_n$
to find the scaled velocity fluctuation $\vert x_n \vert = \delta u_n/\delta u_0$ 
in the range $x_n \sim x_n+dx_n$ can be divided into two parts:
\be
\itPi^{(n)}(x_n) dx_n = \itPi_{\rN}^{(n)}(x_n) dx_n 
+ \itPi_{\rS}^{(n)}(\vert x_n \vert) dx_n
\ee
where the singular part $\itPi_{\rS}^{(n)}(\vert x_n \vert)$, 
stemmed from multifractal distribution of singularities in 
velocity gradient that may be related to the fluctuations caused by
turbulent viscosity, is determined by
$
\itPi_{\rS}^{(n)}(\vert x_n \vert) dx_n = P^{(n)}(\alpha) d \alpha
$
with the transformation of the variables between $\vert x_n \vert$ and $\alpha$:
$
\vert x_n \vert = \delta_n^{\alpha/3}
$,
leading to~\cite{AA4}
\be
\itPi_{\rS}^{(n)}(\vert x_n \vert) dx_n 
= 3G^{(n)}(\vert x_n \vert) \big/\left(Z^{(n)} \vert \ln \delta_n \vert \right) dx_n
\ee
with
\be
G^{(n)}(x) = x^{-1} \left[1 - (\alpha(x) -\alpha_0)^2 \big/(\Delta \alpha)^2
\right]^{n/(1-q)},
\ee
$
\alpha(x) = 3\ln x /\ln \delta_n
$.
Note that the singular part of the probability density function has a dependence
on $\ln \delta u_n$, and that the values of $\vert x_n \vert$ are restricted
within the range $[\delta_n^{\alpha_{\rm max}/3},\ \delta_n^{\alpha_{\rm min}/3}]$.
The first term, $\itPi_{\rN}^{(n)}(x_n)$ assumed to come from thermal or 
dissipative fluctuation caused by the kinematic viscosity, will be considered 
later in this paper after estimating the moments of the velocity fluctuation.

\begin{figure}[htbp]
\begin{center}
\leavevmode
\epsfxsize=70mm
\epsfbox{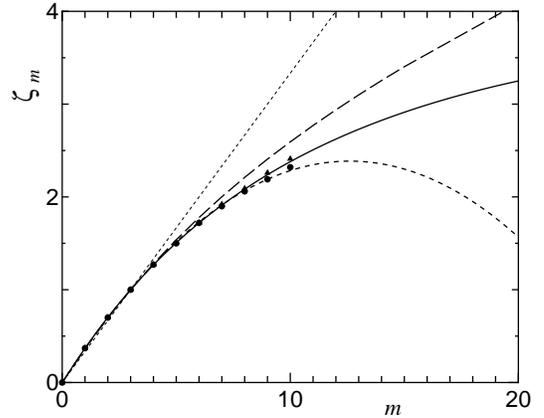}
\caption{The scaling exponent $\zeta_m$ of the velocity structure function.
The experimental results obtained by Lewis and Swinney 
are shown by circles ($\rRe = 540\ 000$) and by triangles ($\rRe = 69\ 000$).
The present theoretical result (\ref{zeta}) is drawn by solid line 
for the intermittency exponent $\mu = 0.28$ taken from the
experimental data.
Dotted line represents K41, whereas dashed line 
She-Leveque.
The prediction of the log-normal model is given 
by short-dashed line with the best fit but inconsistent value $\mu = 0.27$ of 
the intermittency exponent following Lewis and Swinney.
}
\label{Fig: scaling exponent}
\end{center}
\end{figure}

The $m$th moments of the velocity fluctuation, defined by
$
\dbra \vert x_n \vert^m \dket 
= \int_{-\infty}^{\infty} dx_n  
\vert x_n \vert^m \itPi^{(n)}(x_n)
$,
are given by 
\be
\dbra \vert x_n \vert^m \dket = 2 \gamma_m^{(n)} 
+ (1-2\gamma_0^{(n)} ) \
a_m \ \delta_n^{\zeta_m}
\ee
where
$
a_{3\bq} = \{ 2 / [C_{\bq}^{1/2} ( 1+ C_{\bq}^{1/2} ) ] \}^{1/2}
$
with
$
{C}_{\bq}= 1 + 2 \bq^2 (1-q) X \ln 2
\label{cal D}
$,
and
\be
2\gamma_m^{(n)} = \int_{-\infty}^{\infty} dx_n\ 
\vert x_n \vert^m \itPi_\rN^{(n)}(x_n).
\ee
Here, we used the normalization $\dbra 1 \dket = 1$.
The quantity 
\bea
\zeta_m \aeq \alpha_0 m /3 
- 2Xm^2 \big/\left[9 \left(1+{C}_{m/3}^{1/2} \right) \right] 
\nonumber\\
&&- \left[1-\log_2 \left(1+{C}_{m/3}^{1/2} \right) \right] 
\big/(1-q), 
\label{zeta}
\eea
is the so-called scaling exponents of velocity structure function,
whose expression was derived first by the present authors \cite{AA1,AA2,AA3,AA4}.
In Fig.~\ref{Fig: scaling exponent}, we compare the present result 
(\ref{zeta}) with the experimentally measured scaling exponents \cite{Lewis-Swinney99}
at $\rRe = 540\ 000$ (circles) and $\rRe = 69\ 000$ (triangles).
We use the observed value of the intermittency exponent, i.e., 
$\mu = 2 - \zeta_6 = 0.28$, for theoretical analysis.
There are also represented, as references, the predictions of K41 
(dotted line) \cite{K41},  of She-Leveque (dashed line) \cite{She94}
and of the log-normal (short-dashed line) \cite{Oboukhov62,K62,Yaglom}.
Note that our formula (\ref{zeta}) can also explain the scaling exponent 
for higher moments \cite{AA1,AA2,AA3,AA4}.

\begin{figure}[thbp]
\begin{center}
\leavevmode
\epsfxsize=70mm
\epsfbox{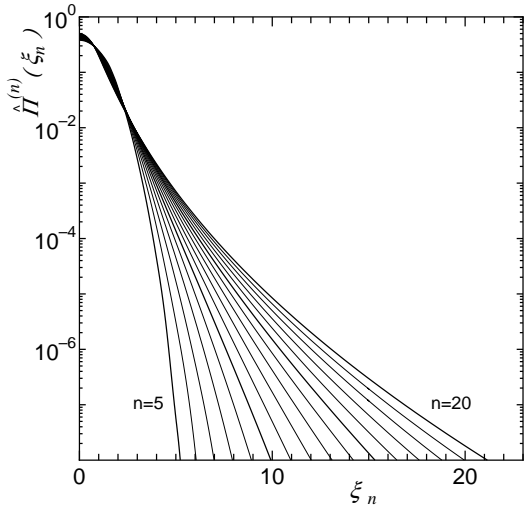}
\caption{The $n$-dependence of the PDF 
$\hat{\itPi}^{(n)}(\xi_n)$ given by the present analysis 
with $q = 0.471$ ($\mu = 0.280$) for integer values $n$ from 5 to 20,
from left to right at the axis $\xi$.}
\label{Fig: Pi-xi theoretical}
\end{center}
\end{figure}

Let us introduce new variable 
\be
\xi_n = \delta u_n / \sqrt{\dbra \delta u_n^2 \dket} 
= x_n / \sqrt{\dbra x_n^2 \dket} = \bar{\xi}_n \delta_n^{\alpha /3 -\zeta_2 /2}
\ee
scaled by the variance of velocity fluctuation where 
$
\bar{\xi}_n = [2 \gamma_2^{(n)} \delta_n^{-\zeta_2} + (1-2\gamma_0^{(n)} ) 
a_2 ]^{-1/2}
$,
and the probability density function (PDF) $\hat{\itPi}^{(n)}(\vert \xi_n \vert)$ 
in terms of this variable through the relation
\be
\hat{\itPi}^{(n)}(\vert \xi_n \vert) d\xi_n 
= \itPi^{(n)}(\vert x_n \vert) dx_n.
\ee
Making use of the relation between $\xi_n$ and $\alpha$, the PDF responsible
for the distribution of singularities 
$\hat{\itPi}_{\rS}^{(n)}(\vert \xi_n \vert)$ can be rewritten in terms of 
$\alpha$ in the form
\be
\hat{\itPi}_{\rS}^{(n)}(\vert \xi_n \vert) = \bar{\itPi}_{\rS}^{(n)}
\delta_n^{\zeta_2/2 -\alpha/3 +1 -f(\alpha)}
\label{Pi-alpha}
\ee
with
$
\bar{\itPi}_{\rS}^{(n)} = 3 (1-2\gamma_0^{(n)} )
/ (2 \bar{\xi}_n \sqrt{2\pi X \vert \ln \delta_n \vert} ).
$

In the following, we will derive the symmetric part of the 
PDF of velocity difference.
The origin of the asymmetry in the PDF (i.e., skewness)
may be attributed to dissipative evolution of eddies or to 
experimental setup and situations.
The consideration of the latter will be reported elsewhere.

\begin{figure}[thbp]
\begin{center}
\leavevmode
\epsfxsize=70mm
\epsfbox{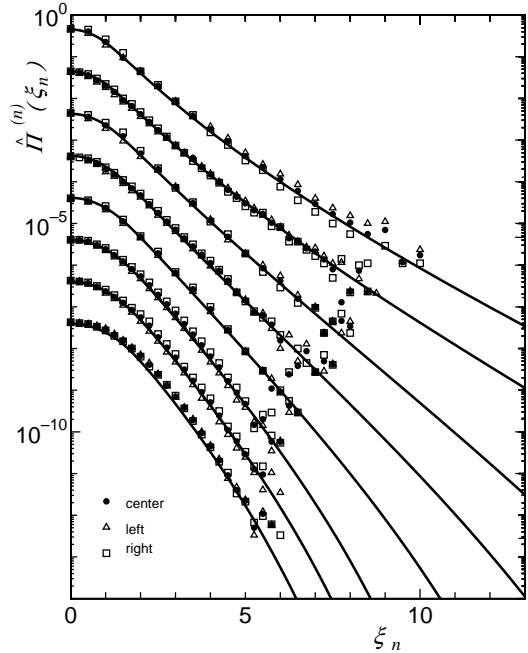}
\caption{Experimentally measured PDF of the velocity fluctuations by 
Lewis and Swinney for $R_\lambda =262$ ($\rRe = 540\ 000$) 
are compared with the present theoretical results $\hat{\itPi}^{(n)}(\xi_n)$.
Open triangles are the experimental data points on the left hand side 
of the PDF taken from Lewis and Swinney, whereas open squares are 
those on the right hand side. 
Closed circles are the symmetrized points obtained by taking 
averages of the left and the right hand experimental data. 
For the experimental data, the distances $r/\eta = \ell_n/\eta$ 
are, from top to bottom: 11.6, 23.1, 46.2, 92.5 208, 399, 830 and 1440. 
Solid lines represent the curves given 
by the present theory with $q = 0.471$.
For the theoretical curves, the number of steps in the cascade $n$ are,
from top to bottom: 14, 13, 11, 10, 9.0, 8.0, 7.5, 7.0.
For better visibility, each PDF is shifted by -1 unit along the vertical axis.
}
\label{experiment by Lewis-Swinney 1999 log}
\end{center}
\end{figure}

In order to determine the normal part $\hat{\itPi}_{\rN}^{(n)}(\xi_n)$ of the PDF, 
we divide 
$\hat{\itPi}^{(n)}(\xi_n)$ into two parts \cite{AA6}:
$
\hat{\itPi}^{(n)}(\xi_n) = \hat{\itPi}_{<*}^{(n)}(\xi_n)
$
for $\vert \xi_n \vert \leq \xi_n^*$, and
$
\hat{\itPi}^{(n)}(\xi_n) = \hat{\itPi}_{*<}^{(n)}(\xi_n)
$
for $\xi_n^* \leq \vert \xi_n \vert \leq 
\bar{\xi}_n \delta_n^{\alpha_{\rm min} /3 -\zeta_2 /2}$.
The point $\xi_n^*$ is defined by 
$
\xi_n^* = \bar{\xi}_n \delta_n^{\alpha^* /3 -\zeta_2 /2}
$
giving 
$
\hat{\itPi}_{\rS}^{(n)}(\vert \xi_n^* \vert) = \bar{\itPi}_{\rS}^{(n)}
$,
where $\alpha^*$ is the solution of 
$
\zeta_2/2 -\alpha/3 +1 -f(\alpha) = 0
$.
The PDF function $\hat{\itPi}_{<*}^{(n)}(\xi_n)$ is constituted 
both by thermal fluctuation and by the 
multifractal distribution of singularities, and is assumed to be given
by Gaussian function \cite{AA6}.
On the other hand, the contribution to 
$\hat{\itPi}_{*<}^{(n)}(\xi_n)$ is assumed to come only from 
the multifractal distribution of singularities, i.e.,
$
\hat{\itPi}_{*<}^{(n)}(\xi_n) = \hat{\itPi}_{\rS}^{(n)}(\vert \xi_n \vert)
$
\cite{AA6}.
Connecting $\hat{\itPi}_{<*}^{(n)}(\xi_n)$ and $\hat{\itPi}_{*<}^{(n)}(\xi_n)$ 
at $\xi_n^*$ having the same value and the same 
derivative there, we have
\be
\hat{\itPi}_{<*}^{(n)}(\xi_n) = \bar{\itPi}_{\rS}^{(n)}
\me^{-[1+3f'(\alpha^*)] [(\xi_n/\xi_n^* )^2 -1 ] /2}.
\ee

The $n$-dependence of $\hat{\itPi}^{(n)}(\xi_n)$ is shown 
in Fig.~\ref{Fig: Pi-xi theoretical}
with the self-consistently determined parameters $\alpha_0$, 
$X$ and $q$ \cite{AA6}.
We notice that there are two points which look like independent of
the number $n$ of steps in the cascade at $\xi_n \approx 0.8$ 
and at $\xi_n \approx 2.3$.
These points are also observed in the experimental data \cite{Lewis-Swinney99}
at about the same values of $\xi_n$.

The comparison of experimentally measured PDF's
of the velocity fluctuation \cite{Lewis-Swinney99} and those obtained by 
the present analysis
is given in Fig.~\ref{experiment by Lewis-Swinney 1999 log}.
In order to extract the symmetrical part from experimental data, we took
mean average of those on the left hand side (represented by
open triangles in the figure) and the right hand side (by open squares).
The symmetrized data are described by closed circles.
The solid lines are the curves $\hat{\itPi}^{(n)}(\xi_n)$ obtained by 
the present analysis.

\begin{figure}[htbp]
\begin{center}
\leavevmode
\epsfxsize=60mm
\epsfbox{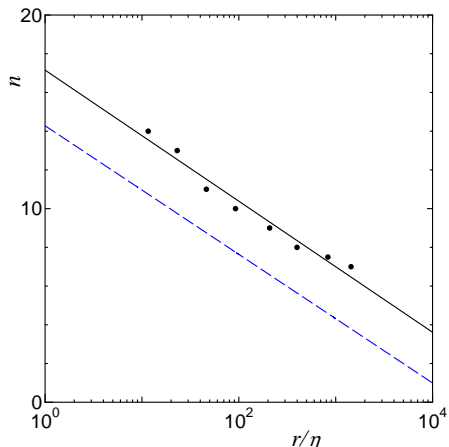}
\caption{Dependence of $n$ on $r/\eta$, extracted from 
Fig.~\ref{experiment by Lewis-Swinney 1999 log}, is given by circles,
which can be fitted by solid line (\ref{n-roeta experimental}).
The dashed line represents (\ref{n-roeta no intermittency}), put for reference, 
which is derived in the absence of intermittency.
}
\label{eoeta-n by Lewis-Swinney 1999}
\end{center}
\end{figure}

The dependence of $n$ on $r/\eta$, extracted from 
Fig.~\ref{experiment by Lewis-Swinney 1999 log}, is shown in 
Fig.~\ref{eoeta-n by Lewis-Swinney 1999} by solid line,
which gives us the best fit
\be
n = -1.019 \times \log_2 r/\eta + 0.901 \times \log_2 \rRe
\label{n-roeta experimental}
\ee
with $\rRe = 540\ 000$.
The dashed line represents (\ref{n-roeta no intermittency}) 
valid in the absence of intermittency.
From (\ref{n-roeta experimental}), we see that the power $w$ in
$
\rRe = (\ell_0/\eta)^{w}
$
turns out to be $w = 1.13$ in the present intermittent turbulence,
which is 4/3 in the absence of intermittency, and that the number
$n_{\rK}$ corresponding to the Kolmogorov scale is estimated 
as $n_{\rK} = 17.2$ for the experiments under consideration.

The success of the present theory in the analysis of 
the experiments \cite{Lewis-Swinney99} may indicate 
the robustness of singularities even for the case of no inertial range.
The same experiments are investigated in \cite{Beck-Lewis-Swinney01}
by a rather different analysis from the present one. 
Comparison of these two approaches is given in \cite{AA6}.
The present theory works quite well also for another systematic 
numerical experiments conducted by Gotoh \cite{Gotoh01}.
It will be reported elsewhere.

The authors would like to thank Prof.~C.~Tsallis for his fruitful comments 
with encouragement.
The authors are grateful to Prof.~T.~Gotoh for his kindness to show his data
prior to publication.


\end{document}